\documentclass[11pt]{article}

\usepackage[margin=1in]{geometry}
\usepackage{graphicx}
\usepackage{tabularx}
\usepackage{booktabs}
\usepackage{xcolor}
\usepackage{amsmath}
\usepackage{amssymb}
\usepackage[hidelinks,colorlinks=false]{hyperref}
\usepackage{xurl}
\usepackage[numbers,sort&compress]{natbib}
\usepackage{caption}
\usepackage{csquotes}
\usepackage[switch]{lineno}

\graphicspath{{figs/}}

\usepackage{tcolorbox}

\newcommand{\toolclass}{agentic command line tool}
\newcommand{\Toolclass}{Agentic command line tool}
\newcommand{\claudecode}{Claude Code}
\newcommand{\copilotcli}{Copilot CLI}

\newcommand{\company}{Microsoft}

\definecolor{initialblue}{HTML}{4682B4}
\definecolor{retentionorange}{HTML}{FF8C00}
\DeclareRobustCommand{\initialmarker}{%
  \tikz[baseline=-0.6ex]{\fill[initialblue] (0,0) circle (0.7ex);}}
\DeclareRobustCommand{\retentionmarker}{%
  \tikz[baseline=-0.6ex]{\fill[retentionorange] (-0.6ex,-0.6ex) rectangle (0.6ex,0.6ex);}}

\definecolor{claudeteal}{HTML}{009E73}
\definecolor{copilotvermillion}{HTML}{D55E00}
\DeclareRobustCommand{\claudemarker}{%
  \tikz[baseline=-0.6ex]{\fill[claudeteal] (-0.7ex,-0.5ex) -- (0.7ex,-0.5ex) -- (0,0.75ex) -- cycle;}}
\DeclareRobustCommand{\copilotmarker}{%
  \tikz[baseline=-0.6ex]{\fill[copilotvermillion] (-0.7ex,0.5ex) -- (0.7ex,0.5ex) -- (0,-0.75ex) -- cycle;}}

\definecolor{careeric}{HTML}{2196F3}
\definecolor{careerm}{HTML}{9C27B0}
\DeclareRobustCommand{\icmarker}{%
  \tikz[baseline=-0.6ex]{\fill[careeric] (0,0) circle (0.7ex);}}
\DeclareRobustCommand{\mmarker}{%
  \tikz[baseline=-0.6ex]{\fill[careerm] (0,0.8ex) -- (0.7ex,0) -- (0,-0.8ex) -- (-0.7ex,0) -- cycle;}}

\newcommand{\RolloutDate}{January 5, 2026}
\newcommand{\CopilotGADate}{February 25, 2026}

\newcommand{\StudyOnePreStart}{October 1, 2025}
\newcommand{\StudyOnePreEnd}{January 4, 2026}
\newcommand{\StudyOnePostStart}{January 5}
\newcommand{\StudyOnePostEnd}{April 29, 2026}
\newcommand{\StudyOnePreWeeks}{13}

\newcommand{\RevPeerInitHigh}{+54}

\newcommand{\SkipPeerInitHigh}{+216}
\newcommand{\SkipPeerRetHigh}{+66}

\newcommand{\ManagerInit}{+82}
\newcommand{\ManagerRet}{+22}

\newcommand{\PriorIdeInitLow}{+49}

\newcommand{\PriorIdeInitHigh}{+83}
\newcommand{\PriorIdeRetLo}{-15}
\newcommand{\PriorIdeRetHi}{-12}

\newcommand{\BasePrInitLow}{+19}

\newcommand{\BasePrInitHigh}{+34}
\newcommand{\BasePrRetLo}{+13}
\newcommand{\BasePrRetHi}{+14}
\newcommand{\BasePrRetHigh}{+31}

\newcommand{\CareerInitJuniorLo}{-14}
\newcommand{\CareerInitJuniorHi}{-13}
\newcommand{\CareerInitSenior}{+22}

\newcommand{\TenureInitNewest}{+11}

\newcommand{\SurveyResponses}{609}

\newcommand{\AbstractPRLift}{24}

\newcommand{\StudyTwoPreStart}{October 1, 2024}
\newcommand{\StudyTwoPreEnd}{January 4, 2026}
\newcommand{\StudyTwoPostStart}{January 5}
\newcommand{\StudyTwoPostEnd}{April 29, 2026}
\newcommand{\PlaceboDate}{2025-10-06}
\newcommand{\StudyTwoPreDays}{461}
\newcommand{\StudyTwoPostDays}{115}

\newcommand{\TreatedWOneWindow}{Jan 5--11, 2026}
\newcommand{\TreatedEverWindow}{Jan 5 -- Apr 29, 2026}

\newcommand{\PersistEarlyLabel}{February}
\newcommand{\PersistLateLabel}{March 1 -- April 29}

\newcommand{\ControlGroups}{10}

\newcommand{\CIHeadlineLift}{+24.0}
\newcommand{\CIHeadlineLiftLo}{+14.5}
\newcommand{\CIHeadlineLiftHi}{+33.7}
\newcommand{\CIHeadlinePText}{$p < 0.001$}

\newcommand{\CIPlaceboLift}{-1.1}
\newcommand{\CIPlaceboLiftLo}{-10.6}
\newcommand{\CIPlaceboLiftHi}{+8.6}

\newcommand{\CIPersistFeb}{+29.4}
\newcommand{\CIPersistFebLo}{+17.7}
\newcommand{\CIPersistFebHi}{+44.4}
\newcommand{\CIPersistMarApr}{+20.0}
\newcommand{\CIPersistMarAprLo}{+7.4}
\newcommand{\CIPersistMarAprHi}{+35.9}

\newcommand{\DoseThreeDay}{+15.0}
\newcommand{\DoseFiveDayPlus}{+50.1}

\newcommand{\ToolClaudeLift}{+11.4}
\newcommand{\ToolClaudeLiftLo}{+9.4}
\newcommand{\ToolClaudeLiftHi}{+13.6}

\newcommand{\ToolCopilotLift}{+24.9}
\newcommand{\ToolCopilotLiftLo}{+23.0}
\newcommand{\ToolCopilotLiftHi}{+26.8}

\newcommand{\ToolContrastPText}{$p < 0.0001$}
\newcommand{\ToolRatio}{2.2}

\newcommand{\CareerIcFourLift}{+21.3}
\newcommand{\CareerIcFourLiftLo}{+19.6}
\newcommand{\CareerIcFourLiftHi}{+23.0}

\newcommand{\TenureFiveFifteenLift}{+21.5}
\newcommand{\TenureFiveFifteenLiftLo}{+19.9}
\newcommand{\TenureFiveFifteenLiftHi}{+23.1}

\title{Adoption and Impact of\\
Command-Line AI Coding Agents\\[2pt]
{\large A Study of \company{}'s
Early 2026 Rollout of \claudecode{} and GitHub~\copilotcli{}}}
\author{
Emerson Murphy-Hill, Jenna Butler, and Alexandra Savelieva\\
Microsoft\\
\texttt{emerson.rex@microsoft.com}\\
\texttt{jennbu@microsoft.com}\\
\texttt{alexandra.savelieva@microsoft.com}
}
\date{}

\begin{document}
\maketitle
\thispagestyle{plain}
\pagestyle{plain}

\begin{abstract}
Organizations rolling out \toolclass{}s like Anthropic's \claudecode{} and
GitHub's \copilotcli{} need to know who will try them, who will keep
using them, and whether the tools produce enough output to justify their cost.
At organizational scale, token spend can run into millions of dollars annually,
so misreading adoption, retention, or impact can make a rollout expensive
without changing engineering velocity. Studying tens of thousands of engineers
at \company{} over its
early-2026 rollout, we find that first use spread primarily through
social networks, retention was associated more with engineers' coding activity
than with demographics, and adopters merged roughly \AbstractPRLift{}\% more pull
requests than they would have otherwise. We use merged pull requests as our proxy
for output --- acknowledging that a merged PR is not the same as the value it
delivers --- and the lift persists across our four-month window.
These results suggest that CLI coding agents are neither
uniformly adopted nor mere novelty effects and that organizations should treat visible
peer use as central to rollout strategy.

\end{abstract}

\section{Introduction}
\label{sec:intro}

\Toolclass{}s like Anthropic's \claudecode{}, Google's Gemini CLI,
and GitHub's \copilotcli{} are increasing in popularity among software developers.
Such tools harness agents that call large language models, where the agents
execute semi-autonomous commands on behalf of the user from the command line. In
early 2026, the Pragmatic Engineer's survey indicated that
\claudecode{} was the most popular AI-based developer
tool among respondents~\citep{pragmaticengineer2026}.

At the same time, organizations considering whether to purchase such tools
do not yet know which of their engineers are likely to adopt and use them.
At the end of 2025, StackOverflow's
survey~\citep{stackoverflow2025} of more than 49,000 respondents indicated that
developers' trust in AI is falling and that ``developers remain willing but
reluctant to use AI''.

Even \emph{if} organizations choose to use AI, the tokens needed to execute
these tools can be expensive. At the extreme high end, Fortune
reports on Meta employees' usage of AI~\citep{fortune2026meta}:

\begin{quote}
\emph{In a 30-day period, total employee usage on the dashboard exceeded 60
trillion tokens, and the highest-ranked individual user averaged 281 billion
tokens. Using the least expensive version of Claude Opus 4.6, which costs
\$5 for every million tokens, that one user alone
could have cost Meta more than \$1.4 million.}
\end{quote}

\noindent
Even at more modest levels of usage, organizations may wonder what
return on investment they should expect.

Three concerns thus follow any rollout: \emph{which engineers will adopt},
\emph{whether they will keep using the tool}, and \emph{whether the tool
produces enough additional output --- which we operationalize as merged pull
requests --- to justify its cost}.
To examine them, we analyze our
experience in early 2026 with \company{} offering its engineers two
\toolclass{}s --- \claudecode{} and \copilotcli{} --- over a roughly four-month
window of usage and pull request (PR) activity.

This paper contributes the first field study to use developer-level telemetry
to analyze both the adoption of \toolclass{}s and their effect on pull-request output.
Prior developer AI adoption work typically relies on surveys and interviews,
and prior impact work largely infers AI use from public-repository signals (Section~\ref{sec:related-work});
our enterprise setting instead
observes every engineer who could adopt alongside direct usage. Within this
setting we separate initial use from retention, trace merged-PR output against
how intensively the tools are used, compare two different tools, and draw on an
internal developer survey to help interpret the results.

We analyze adoption and impact in two studies. The \emph{adoption study}
(Section~\ref{sec:study1}) asks who adopts: among engineers who could use
\copilotcli{}, who tries it (RQ1) and, among those who try it, who keeps
using it (RQ2). The \emph{outcomes study} (Section~\ref{sec:study2}) asks what adoption
produces: whether using either tool yields more merged PRs (RQ3), whether the
specific tool matters (RQ4), and which engineers benefit most (RQ5). The adoption
study covers only \copilotcli{}, the tool with a well-defined eligible-adopter
population at rollout; the outcomes study covers both \claudecode{} and \copilotcli{}.

\section{Related Work}
\label{sec:related-work}

\subsection{Adoption of Developer AI Assistance}

The study of technology adoption has a long history in the social
sciences, from Rogers' diffusion of innovations theory~\citep{rogers2003diffusion}
to more technologically focused work on information systems~\citep{davis1989perceived,venkatesh2000theoretical,venkatesh2003user},
which tells us that adoption is shaped by individual, social, and organizational
factors. In this paper, we use \emph{adoption} as an umbrella term for a developer's uptake of
a tool, and decompose it into two phases that this literature treats as
distinct~\citep{rogers2003diffusion,bhattacherjee2001continuance}: \emph{initial
use} --- a developer's first use --- and \emph{retention} --- whether
that use is sustained. We study the two separately because, as our results show,
the factors that predict initial use are not those that predict retention.

In the developer AI space, Reyes-Reina and colleagues' recent
systematic review of AI-tool adoption in software development characterized 25
studies~\cite{reyes2026adoption}.
Yet every empirical study in that review rests on
surveys, interviews, or focus groups --- or, where behavior is directly observed,
on one-shot lab tasks; \emph{none} draw on observational data of developers'
actual adoption. We fill exactly this gap, enabled by access to
(1) longitudinal, individually-identifiable telemetry of AI tool usage and
(2) human resources (HR) data about developers.

Beyond that review, a few recent studies examine developer AI adoption with
observational data.
Daniotti and colleagues infer AI-assisted coding from a commit classifier
 to document AI's global diffusion~\cite{daniotti2026using};
Yang identifies \claudecode{} adopters among 16,000 scientists via the tool's
default commit co-author trailer~\cite{yang6803624claude};
and Robbes and colleagues detect AI agent adoption across GitHub projects from
commit trailers, config files, and branch names~\cite{robbes2026agentic}.
But all observe only developers who leave a public-repository signal, so a
``non-adopter'' is merely someone with no such signal --- conflating true non-use
with suppressed or invisible use.
Lacking a roster of \emph{eligible} engineers, they cannot cleanly separate adopters
from non-adopters.
Our enterprise setting supplies that missing denominator: the full population of
engineers who \emph{could} adopt \company{}'s sanctioned \toolclass{}s, paired
with logs of who tries AI and who keeps using it.

\subsection{Impact of Developer AI Assistance}

How AI assistance affects developer productivity has been examined along three
axes: what developers \emph{report}, how they perform on
\emph{controlled tasks}, and what they \emph{ship} in real-world work.

The most common evidence of developer productivity is self-reported.
A recent systematic review finds that surveys of developer satisfaction are the
dominant way studies measure productivity~\cite{mohamed2026}.
Such surveys generally find that developers who accept more AI
suggestions report feeling more productive and
satisfied~\cite{ziegler2022productivity,o2025more,afroz2026}, and they point to
writing and implementing code as where AI helps most~\cite{gurgul2026state}.
What developers report, however, need not match what they ultimately ship.

A second line of evidence comes from controlled experiments, which can establish
causation but vary in how well their tasks resemble real work.
In such studies, results are mostly but not uniformly positive:
several studies find faster task completion with
AI~\cite{peng2023,weber2024significant,shihab2025effects,brandebusemeyer2025developers,paradis2025much},
while a few report no significant difference or considerable time spent verifying
AI output~\cite{vaithilingam2022expectation,mozannar2024reading}.
Studies that instead let developers bring their own work improve
ecological validity but shrink in scale: AI resolved only about half of
developers' real tasks in one study of 19 participants~\cite{kumar2025} and
\emph{increased} completion time by 19\% in another study of 16 developers~\cite{becker2025measuring}.
Field experiments that randomly assign AI to developers in their natural environment
are likewise split, from no significant change in merged pull
requests~\cite{Cui2024Productivity,butler2025} to a significant
pull request lift~\cite{cui2025}.

The third line of evidence, closest to our own, observes developers'
output in real-world settings. Some studies compare output before and after a
cohort adopts AI and find little movement: 39 developers at one company showed no
change in commit counts over two years~\cite{stray2026}, and three agile teams
showed no change in commit volume~\cite{tomaz2026impacts}, though such
before-and-after comparisons are vulnerable to seasonal confounds.
Others infer AI use from indirect signals --- committed configuration files, commit co-author
trailers, or initially supported languages --- and report effects ranging from
large initial lifts that fade after a few
months~\cite{he2026cursor,quispe2026coding,agarwal2026agentic}, to no effect~\cite{kreitmeir2024heterogeneous},
to up to 40\% increases in merged pull requests~\cite{yeverechyahu2024impact}.
Inferring use from indirect signals, however, admits both false positives and false negatives
in terms of actual AI usage.

The studies closest to ours sidestep that problem with subscription or telemetry data that identify AI
use directly, reporting merged-PR lifts that range from a few percent to over
60\%~\cite{Demirer2026,song2024impact,kumar2025intuition,heilman2026ghcp}.
However, none of these studies use telemetry for \toolclass{}s specifically --- they instead
largely study IDE-based AI tools.
This is the gap our outcome study fills,
measuring how usage of \toolclass{}s affects pull request throughput in a real-world setting.

\section{Background}
\label{sec:background}

In the first half of 2026, \company{} engineers had access to two sanctioned
\toolclass{}s: \copilotcli{} and \claudecode{}. GitHub announced
general availability of \copilotcli{} on \CopilotGADate{}; \company{} had
earlier access through a product preview program. \claudecode{} reached \company{} engineers along a
narrower path: starting in late 2025, specific divisions and individuals
received access through a managed program, and the access list evolved as the
program matured. We refer throughout to \RolloutDate{} --- the boundary
between our pre-period and post-period, and the point around which \company{}'s
broader internal use of these tools took off --- as \emph{the rollout}. This
rollout asymmetry shapes the adoption study's sample; the
outcomes study, which examines engineers who did adopt, covers both
tools.
Both studies' observation windows close on \StudyTwoPostEnd{}; shortly afterward,
an internal announcement indicated that \claudecode{} licenses would be
discontinued for most engineers in about a month, with affected engineers directed
to transition to \copilotcli{}. We end our analysis window before this point because including weeks
in which engineers were being moved off \claudecode{} would muddle our results:
migrating engineers would look like organic \copilotcli{} adopters, and \claudecode{}
usage would fall for policy reasons rather than engineer preference.

Prior to this period, employees at \company{} already had access to
GitHub Copilot in non-CLI forms for several years.
These included Copilot for code completion, chat, and agent mode in IDEs like
VS Code.
Thus, readers should not interpret our findings broadly as being about AI tools,
but rather as about the adoption and impact of \toolclass{}s specifically
in a context where non-CLI tools were already available.

\section{The Adoption Study}
\label{sec:study1}

The adoption study examines \copilotcli{} uptake in two phases --- initial use
and retention --- among the engineers eligible to adopt it at rollout, the sample
we define in Section~\ref{sec:study1-sample}. We also draw on a developer survey
(Section~\ref{sec:qual-methods}) to help interpret the adoption patterns we find.

The quantitative pipelines and figures in both studies were implemented in code
written with an AI assistant under the authors' direction and verified by the
authors; we give our full AI-use disclosure in the Acknowledgments.

\subsection{Research questions and outcome variables}
\label{sec:study1-outcomes}

The adoption study answers two questions that, although they sit naturally
together, require different data structures and different statistical models. We state
each question with the outcome variable that operationalizes it.

\textbf{RQ1. Among engineers who could have used \copilotcli{} but had not yet,
who tries it first?} We define initial use as an engineer's
first opening \copilotcli{} during the post-rollout window.
For each engineer-week $(i,w)$ in which engineer
$i$ had no prior \copilotcli{} activity, the initial-use outcome $A_{i,w}$ is 1
if $i$ recorded any \copilotcli{} activity in week $w$ and 0 otherwise; once
$A_{i,w} = 1$, engineer $i$ contributes no further rows.
Because the population still at risk of first use shrinks from
week to week, this gives a discrete-time hazard on an engineer-week panel.

\textbf{RQ2. Among engineers who adopted \copilotcli{}, who keeps using it?} We
define retention as sustained early use: an adopter is retained if
they recorded \copilotcli{} activity on at least 5 of the 14 days beginning with
their first use. The 5-of-14 threshold approximates ``used \copilotcli{} on
roughly half of working days during the first two weeks,'' designed to separate
``tried and stayed'' from ``tried and abandoned.'' We exclude adopters whose
first use was within 14 days of the end of our tool-use observation window,
since their retention window has not yet fully elapsed. Because retention is
defined only for engineers who have already adopted, it is a cross-sectional
outcome with one row per adopter.

\subsection{Sample}
\label{sec:study1-sample}

The adoption study's sample consists of \company{} software engineers, every one
of whom could have tried \copilotcli{} on the rollout date. Engineers who already had
\claudecode{} access, by contrast, faced a different
adoption decision, with a sanctioned \toolclass{} alternative already on hand. To
isolate the population whose adoption decision is uniform, we start from all
\company{} employees with software engineer job titles and drop three groups:

\begin{itemize}
  \item members of two large divisions that received broad \claudecode{} access
  through a managed program;
  \item individual \claudecode{} licensees outside those divisions;
  \item \emph{retracted} users --- engineers who once held a \claudecode{}
  license but no longer do.
\end{itemize}

\noindent
Since \copilotcli{} was generally available from GitHub starting
\CopilotGADate{} but was available to \company{} through a product preview program somewhat before that,
we chose \RolloutDate{}, an earlier date consistent with our outcomes study, as
the cutoff.
We thus excluded engineers who first used \copilotcli{} prior to this cutoff.

The pre-period for predictor construction runs \StudyOnePreStart{} through
\StudyOnePreEnd{} (\StudyOnePreWeeks{} weeks). The post-period over which adoption and
retention are observed runs \StudyOnePostStart{} through \StudyOnePostEnd{}.

\subsection{Predictors}
\label{sec:study1-predictors}

We collected five predictor groups for each engineer. For each group we name
the construct, why it might shape initial use or retention, the operationalization,
and the source.

\paragraph{Career stage} The construct is the scope, qualifications, and
general expectations of the engineer's role. \company{} uses parallel
career-stage ladders --- IC for individual contributors and M for
people-managers --- where higher numbers indicate greater scope and seniority
within each ladder, and stages at the same number on the two ladders (e.g.,
IC4 and M4) are calibrated as peers. Operationalization: career stage from
HR records, coded as IC2--IC6 (individual contributors) and M4--M6 (managers), with IC4 as the reference category.

\paragraph{Tenure} The construct is years at \company{}. Operationalization: years since
first hire date, bucketed into $<$1y, 1--2y, 2--5y, 5--15y (the largest bucket,
used as reference), and 15+y.

\paragraph{Baseline pull-request activity} The construct is how actively the
engineer was merging code before the rollout. More active engineers have more
occasions on which an \toolclass{} could plausibly help. Operationalization: mean
PRs created per week over the \StudyOnePreWeeks{}-week pre-period, bucketed as 0 (reference),
$\leq$1, 1--2, and 2+.

\paragraph{Prior IDE Copilot use} The construct is openness to AI tooling
already demonstrated in a related product. Prior AI tool users may be quicker
to try a new AI tool; alternatively, they may be sufficiently satisfied with
what they have that \copilotcli{} offers little marginal value.
Operationalization: active days using IDE Copilot over the \StudyOnePreWeeks{}-week
pre-period, bucketed as 0 (reference), 1--14, 15--60, and 60+. We count a day
as an active day if the engineer accepted at least one autocomplete suggestion
or sent at least one chat message in IDE Copilot on that day.

\paragraph{Social exposure} The construct is whether the colleagues in an
engineer's professional orbit --- the people they collaborate with, share a
reporting chain with, or report to --- are visibly using \copilotcli{}. We compute
three time-varying signals for each engineer-week over the
prior 14 days:
\begin{itemize}
  \item \emph{Reviewer peers}: the share of engineers with whom engineer $i$ exchanged
  at least three reviews during the pre-period who used \copilotcli{}.
  \item \emph{Skip-level peers}: the share of engineers who report up to the same
  manager's manager as engineer $i$ and who used \copilotcli{}.
  \item \emph{Direct manager}: a binary indicator that $i$'s direct manager
  used \copilotcli{}.
\end{itemize}
The reviewer-peer and skip-level-peer shares are bucketed as 0\% (reference),
1--10\%, 10--25\%, and 25+\%; the manager indicator is binary.

\subsection{Controls}
\label{sec:study1-confounds}

Three classes of confounder threaten the predictor estimates, and we absorb each
in our model specification.

\begin{itemize}
  \item \textbf{Calendar-time effects.} Rollout momentum, weekly seasonality, and
  holidays can shift initial use for everyone at once; week fixed effects
  identify each coefficient within calendar week.
  \item \textbf{Division-level effects.} Some divisions may adopt earlier as a
  whole --- through leadership encouragement or differential exposure to internal
  communications --- so broad-division fixed effects identify each coefficient
  within division.
  \item \textbf{Within-engineer repeated observations.} An engineer contributes
  one row per at-risk week, so the rows are not independent; we cluster standard
  errors on engineer.
\end{itemize}

\subsection{Model specifications}
\label{sec:study1-models}

\paragraph{Initial-use model} We fit a discrete-time logistic regression on the
pre--first-use engineer-week panel:
\begin{equation}
  \mathrm{logit}\,\Pr\!\left[A_{i,w} = 1\right]
  = \alpha_w + \gamma_{o(i)} + \beta^{\top} X_i + \delta^{\top} S_{i,w},
  \label{eq:adoption}
\end{equation}
where $\alpha_w$ are week dummies, $\gamma_{o(i)}$ are broad-division dummies
for $i$'s division $o(i)$, $X_i$ is the vector of pre-rollout predictors
(career stage, tenure, baseline PR activity, prior IDE Copilot use), and
$S_{i,w}$ is the vector of time-varying social-exposure signals.

\paragraph{Retention model} We fit a cross-sectional logistic regression on
the adopters:
\begin{equation}
  \mathrm{logit}\,\Pr\!\left[R_i = 1\right]
  = \alpha_{w^*(i)} + \gamma_{o(i)} + \beta^{\top} X_i + \delta^{\top} S_{i,\,w^*(i)},
  \label{eq:retention}
\end{equation}
where $R_i = 1$ if engineer $i$ was active on at least 5 of the 14 days from
first use, $w^*(i)$ is $i$'s first-use week, and the time-varying social
predictors are evaluated at $w^*(i)$.

We  report each predictor as the
percent change in odds compared to its reference category, with 95\%
confidence intervals.

\subsection{Qualitative Data}
\label{sec:qual-methods}

To contextualize the telemetry, we surveyed attendees of \company{}'s internal
``Agentic Engineering Day'' --- a set of demonstrations, tutorials, and hands-on
sessions on these tools --- two weeks after the event, with open-ended questions
about how their work had changed: its productivity, its nature, how saved time
was reallocated, and whether the tools shifted work toward higher-value tasks or
demanded more oversight. From the \SurveyResponses{} responses, an author
identified recurring themes and selected representative quotes that help
interpret the quantitative findings.

\subsection{Results}
\label{sec:adopt-results}

We organize the results around five predictor groups, presented in order of how
strongly they are associated with initial use: social exposure, prior IDE
Copilot use, baseline pull-request activity, career stage,
and tenure. Each group has one figure showing how that group's
predictors shift the odds of initial use (blue circles) and retention (orange
squares) compared to the reference engineer. Throughout, filled markers are
statistically significant at $p<0.05$ and open markers are not; each point
estimate is printed at the right of its row.

\subsubsection{Social exposure: peers and managers}
\label{sec:adopt-r-social}

\begin{figure}[!htbp]
  \centering
  \includegraphics[width=0.75\columnwidth]{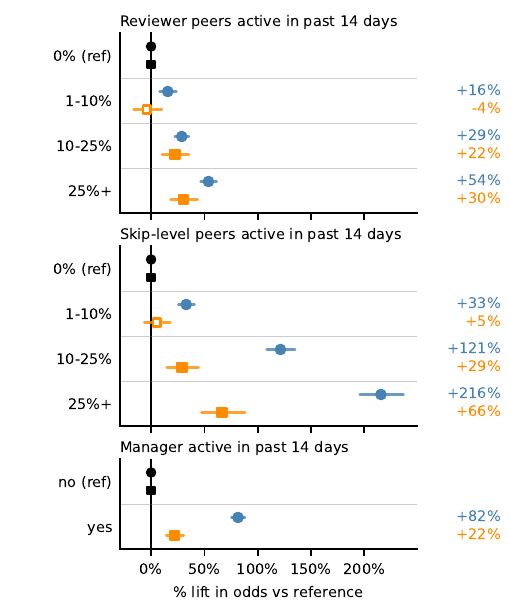}
  \caption{Change in odds of initial use~\initialmarker{} and
  retention~\retentionmarker{} of \copilotcli{} by social exposure, versus an
  engineer with no coworkers who used \copilotcli{} in the prior 14 days.}
  \label{fig:adopt-social}
\end{figure}

Figure~\ref{fig:adopt-social} breaks social exposure into three groups of
coworkers. The pattern is consistent across all three: an
engineer whose coworkers were already using \copilotcli{}
had higher odds of trying it themselves. Each row compares a given level of
coworker use against no coworker use:

\begin{itemize}
  \item \textbf{Reviewer peers --- the engineers someone regularly trades code
  reviews with.} The more of them were already using \copilotcli{}, the more likely
  the engineer was to try it: odds reached \RevPeerInitHigh{}\% higher where a
  quarter or more had used it, with retention rising similarly.
  \item \textbf{Skip-level peers --- the engineers who share a skip-level
  manager.} This was the strongest signal: an engineer whose skip-level peers were
  largely using \copilotcli{} had \SkipPeerInitHigh{}\% higher odds of trying it
  where more than a quarter had. The association with retention was more modest
  (\SkipPeerRetHigh{}\% at the top level).
  \item \textbf{Direct manager.} An engineer whose manager used \copilotcli{} had
  higher odds of both trying it (\ManagerInit{}\%) and, more modestly, sticking
  with it (\ManagerRet{}\%).
\end{itemize}

The survey hints at why peer usage matters. Developers said the tools let them
contribute more broadly to team activities: one described ``updating our
documentation, analyzing it for issues and quality, prototyping app ideas and
code samples, creating tools for our team and others to use.'' Another connected
this breadth to saved time, saying, ``I do save time [...] I am able to utilize
my time in coming up with numerous ideas that are very useful for my team.'' Such
shared output---tools, docs, code---may benefit the wider team and reinforce peer
uptake.

\subsubsection{Prior IDE Copilot Use}
\label{sec:adopt-r-ide}

\begin{figure}[!htbp]
  \centering
  \includegraphics[width=0.75\columnwidth]{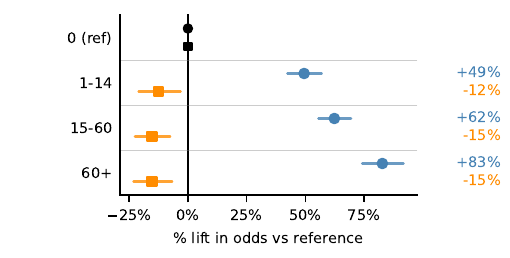}
  \caption{Change in odds of initial use~\initialmarker{} and retention~\retentionmarker{} of \copilotcli{} by prior IDE Copilot use, versus an engineer who did not use IDE Copilot during the pre-period.}
  \label{fig:adopt-ide}
\end{figure}

Figure~\ref{fig:adopt-ide} groups engineers by how many days they used IDE
Copilot during the \StudyOnePreWeeks{}-week pre-period, from none (the reference) up to
60+ days.
Unlike the social signals, prior IDE use cut in opposite directions for the two
outcomes:

\begin{itemize}
  \item \textbf{Engineers who had used IDE Copilot were more likely to try
  \copilotcli{}.} The more days of prior IDE use, the higher the odds of trying
  \copilotcli{} --- from \PriorIdeInitLow{}\% higher at 1--14 days up to
  \PriorIdeInitHigh{}\% at 60+ days.
  \item \textbf{But those same engineers were less likely to stick with
  \copilotcli{}.} All three retention markers are negative (between $\PriorIdeRetHi$\%
  and $\PriorIdeRetLo$\%): an engineer who leaned on IDE Copilot before the rollout
  retained \copilotcli{} at a lower rate than one who had never used it.
\end{itemize}

\subsubsection{Baseline pull-request activity}
\label{sec:adopt-r-prs}

\begin{figure}[!htbp]
  \centering
  \includegraphics[width=0.75\columnwidth]{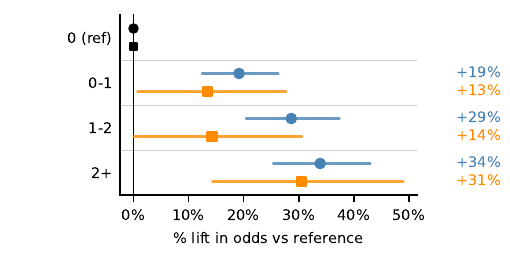}
  \caption{Change in odds of initial use~\initialmarker{} and retention~\retentionmarker{} of \copilotcli{} by baseline pull-request activity, versus an engineer who created no pull requests during the pre-period.}
  \label{fig:adopt-prs}
\end{figure}

Figure~\ref{fig:adopt-prs} groups engineers by how many pull requests they
created per week during the \StudyOnePreWeeks{}-week pre-period --- a proxy for
how actively they were already shipping code --- from none (the reference) up to
2 or more.
More baseline activity went with trying \copilotcli{}, but only
the busiest engineers were also more likely to stay:

\begin{itemize}
  \item \textbf{Engineers who were already shipping code were more likely to try
  \copilotcli{}.} The more pull requests an engineer created per week, the higher
  the odds of trying \copilotcli{} --- from \BasePrInitLow{}\% higher at up to 1
  PR/week to \BasePrInitHigh{}\% at 2 or more.
  \item \textbf{Busier engineers were also more likely to stay, increasingly so
  the more they shipped.} Retention rose with prior PR volume at every level
  --- \BasePrRetLo{}\% up to 1 PR/week, \BasePrRetHi{}\% at 1--2, and
  \BasePrRetHigh{}\% at 2 or more.
  The retention advantage is more pronounced at the highest activity level.
\end{itemize}

\subsubsection{Career stage}
\label{sec:adopt-r-career}

\begin{figure}[!htbp]
  \centering
  \includegraphics[width=0.75\columnwidth]{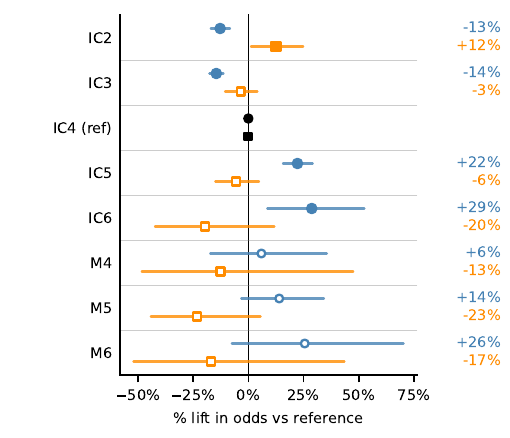}
  \caption{Change in odds of initial use~\initialmarker{} and retention~\retentionmarker{} of \copilotcli{} by career stage, versus a mid-level individual contributor (IC4).}
  \label{fig:adopt-career}
\end{figure}

Figure~\ref{fig:adopt-career} groups engineers by their career stage from HR
records, sorted from the most junior individual contributors (IC2) through senior
ICs to managers (M4--M6), each compared against a mid-level individual
contributor (IC4). Among ICs there was a gentle seniority gradient in trying
\copilotcli{}, while managers looked no different from the reference:

\begin{itemize}
  \item \textbf{Junior ICs were less likely to try \copilotcli{}.} IC2 and IC3
  engineers had lower odds of trying it than a mid-level IC ($\CareerInitJuniorHi$\%
  and $\CareerInitJuniorLo$\%); their retention
  markers were noisy, with only IC2's statistically significant.
  \item \textbf{Senior ICs were more likely to try it.} IC5 and IC6 engineers had
  higher odds than a mid-level IC --- about \CareerInitSenior{}\% for IC5 --- but
  their retention markers sat near zero.
  \item \textbf{Managers looked no different from the reference.} M4--M6 engineers
  showed no statistically distinguishable difference from IC4 on either trying
  \copilotcli{} or sticking with it; the point estimates wander slightly but their
  confidence intervals are wide.
\end{itemize}

\begin{figure}[!htbp]
  \centering
  \includegraphics[width=0.75\columnwidth]{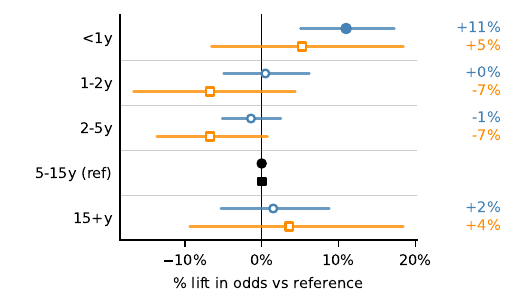}
  \caption{Change in odds of initial use~\initialmarker{} and
           retention~\retentionmarker{} of \copilotcli{} by tenure,
           versus an engineer who had been at \company{} for 5--15 years.}
\label{fig:adopt-tenure}
\end{figure}

The survey helps explain these seniority differences. One developer called the tools
``an especially good fit for experienced, senior developers, who can break work
down into smaller chunks that the AI [...] can implement[...] and then
review[...]'' for ``correctness, completeness, and sanity.'' The same respondent
added, ``I'm less convinced that junior developers (who `don't know what they
don't know') would be able to use them as effectively.'' Seniors also carry work
beyond coding---architecture, review, coordination---and several said the tools
let them offload coding while staying in those roles. One principal-level
developer said coding assignments often required 5--10 clarification prompts and
several more ``I see this error'' prompts, with each iteration taking 10--15
minutes. Even so, that latency was useful because ``now I can just prompt it and
switch to another task (reviewing PRs, reviewing design docs or just meetings).''

Others worried about juniors specifically, which may help explain their lower
initial use---one flagged ``what these tools mean for junior colleagues and how
they can develop a good `sense' for code [...] to know when [the] output is not
optima[l]'' or how to ``steer it into the best solution for the task at hand.''
Together these responses suggest seniors are better positioned to exploit
agentic tools---decomposing tasks, vetting outputs, and juggling parallel
work---which dovetails with their modest edge in \emph{trying} \copilotcli{}; that
same spare capacity may feed shared team resources, reinforcing the peer-usage
association.

\subsubsection{Tenure}
\label{sec:adopt-r-tenure}

Figure~\ref{fig:adopt-tenure} groups engineers by how many years they had been
at \company{} --- from under a year up to 15 or more --- compared against the
5--15-year bucket, the largest group and the workforce midpoint. Tenure barely
mattered:

\begin{itemize}
  \item \textbf{The newest hires were a little more likely to try \copilotcli{}.}
  Engineers in their first year had \TenureInitNewest{}\% higher odds of trying it
  than the reference.
  \item \textbf{Every other group looked similar.} The 1--2y, 2--5y,
  and 15+y buckets all sat within a couple of percent of the reference on initial
  use, and none of their markers --- initial use or retention --- were
  statistically significant.
\end{itemize}

\section{The Outcomes Study}
\label{sec:study2}

The outcomes study turns from who adopts, to what adoption produces. Unlike the
adoption study, which examined only \copilotcli{}, it covers both \copilotcli{}
and \claudecode{}. Because the outcomes study conditions on engineers who had
already adopted --- rather than modeling the adoption decision itself --- it is
not constrained by the rollout asymmetry that limited the adoption sample to
\copilotcli{}, so both tools can be studied together.
We say ``engineers'' throughout for brevity; the outcomes cohort is defined by tool use
and pull-request activity rather than job title, so it may include other
development roles doing engineering-like work.
The outcome throughout is merged-code throughput --- the count of \emph{merged} pull
requests, counting a PR as merged if it completes within 28 days of creation.
This window keeps the outcome comparable across the post-period and avoids
right-censoring PRs created near the data cutoff; we discuss its
construct-validity implications in Section~\ref{sec:threats}.

We pursue three questions across two self-contained parts, each with its own
dataset, design, results, and limitations:

\textbf{RQ3. Does using \toolclass{}s lead to more merged PRs?} \textbf{Part~1}
estimates the average lift for early adopters from a
synthetic-control counterfactual built with CausalImpact; \textbf{Part~2}
revisits it with a within-person dose-response that makes
each engineer their own control across weeks of differing use.

\textbf{RQ4. Does the specific tool matter --- is \claudecode{} different from
\copilotcli{}?} Part~2 compares the tools among engineers who
used only one.

\textbf{RQ5. Among engineers who use these tools, who benefits most?} Part~2
interacts the dose with career stage and tenure.

\begin{figure}[!htbp]
  \centering
  \includegraphics[width=\columnwidth]{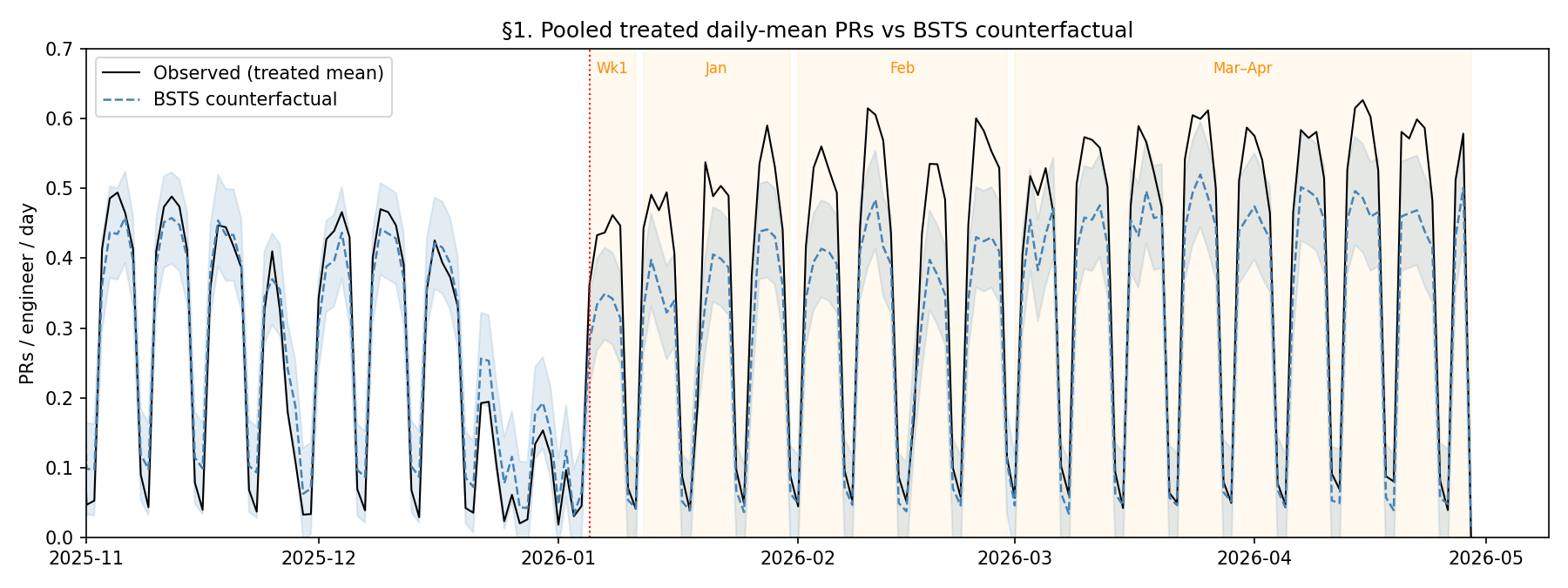}
  \caption{Daily merged PRs per engineer for single-tool adopters, observed versus
  their synthetic counterfactual.}
  \label{fig:headline}
\end{figure}

\subsection{Part 1: Synthetic-control estimate (CausalImpact)}
\label{sec:rq3}
\label{sec:rq3-headline}

\paragraph{Dataset}
Part~1 uses a rollout-aligned adopter cohort and a
non-adopter control, with the outcome aggregated as merged PRs per engineer per
day. The dataset comprises adopters (\claudecode{} or \copilotcli{},
including mixed-tool users) with any tool activity during \TreatedWOneWindow{},
forming a rollout-aligned cohort. We fix the cohort to this single
intervention week because the CausalImpact design compares one pre-period
trajectory to one post-period trajectory. The control is a sample of PR creators
who never used either tool, constructed as described next. Both cohorts are
restricted to active engineers --- at least two merged PRs in a four-week
pre-rollout window --- so that the estimate is not diluted by engineers doing
little or no PR-based work. We apply a P5--P95 PR
filter on pre-period PR counts to trim the tails that destabilize the estimates;
the estimate is directionally unchanged when this filter is dropped.
The pre-period runs \StudyTwoPreStart{} -- \StudyTwoPreEnd{} (\StudyTwoPreDays{} days)
and the post-period \StudyTwoPostStart{} -- \StudyTwoPostEnd{} (\StudyTwoPostDays{} days).

\paragraph{Design}
We use the Bayesian Structural Time-Series (BSTS) implementation of
CausalImpact~\citep{brodersen2015causalimpact}.
The synthetic counterfactual is built from a pool of \ControlGroups{} daily-mean
regressors: we sampled PR creators who never used either tool,
randomly partitioned them into \ControlGroups{} groups, and averaged each
group's daily PR-per-engineer series.
The
inference quantity is the posterior mean lift over the \StudyTwoPostDays{}-day post-period
with a 95\% credible interval, accompanied by a posterior tail-area
$p$-value.

The design absorbs any factor that moves adopters and non-adopters together
(aggregate PR-volume trends, weekly seasonality, company-wide events). It does
\emph{not} absorb adopter-specific shocks --- for example, if adopters
disproportionately work on a product that shipped during the post-period; we probe
for such shocks with the placebo test described in Section~\ref{sec:threats}.

\paragraph{Results}
Figure~\ref{fig:headline} plots the observed merged-PR series for adopters against the
counterfactual of how many they would have merged had they not adopted the
\toolclass{}s. The observed series is solid
black; the model's estimate of that counterfactual is the dashed blue line, with its
95\% credible interval shaded, and the gap between the two is the lift attributable to
adoption. The vertical line marks the rollout date.

CausalImpact estimates a \textbf{\CIHeadlineLift{}\%} lift in PRs/engineer/day over the
post-period [95\% CI \textbf{\CIHeadlineLiftLo{}\%, \CIHeadlineLiftHi{}\%}], posterior tail-area
\CIHeadlinePText{}.

Survey responses echo this lift directly, with developers noting
``I complete more PRs per week'', ``This has freed me up to do more PRs'', and
``I am able to make PRs faster and understand the codebase easily''.

\paragraph{Persistence: does the lift fade?}
\label{sec:rq3-persistence}

He and colleagues~\citep{he2026cursor} showed that an early productivity lift
from adopting Cursor in open source --- measured in commits and lines of code ---
faded within a few months. Is the same true of our \toolclass{}s at \company{}
in 2026?

Our data indicate ``no''. Figure~\ref{fig:headline} splits the post-period
into four successive buckets --- the first week, the rest of January, February,
and March--April. Comparing the two later monthly buckets, the lift
shows no statistically distinguishable decay:
\textbf{\CIPersistFeb{}\%} (95\% CI [\CIPersistFebLo{}\%, \CIPersistFebHi{}\%]) in
\PersistEarlyLabel{} versus \textbf{\CIPersistMarApr{}\%} (95\% CI [\CIPersistMarAprLo{}\%,
\CIPersistMarAprHi{}\%]) across \PersistLateLabel{}. These posterior intervals overlap
substantially and both exclude zero, so the drop in the point estimate is within
sampling noise; within the resolution of these buckets, the productivity gain is
sustained, not transient.

The survey suggests this persistence is more than novelty---it points to durable
shifts in how developers work. As one put it, ``Using GitHub CLI has entirely
changed the way that I approach all my projects. I no longer think about narrow
solutions; instead I am able to use agents to think broadly and formulate
wholistic approaches. I feel so much more productive, I am never going back.''

\subsection{Part 2: Within-person analyses}
\label{sec:study2-within}

Part~2 holds each engineer as their own control, comparing weeks in which they
used a tool to weeks in which they did not. This within-person design addresses
three limitations of Part~1: it lets the estimated benefit vary across
engineers and across levels of use, rather than collapsing everyone into one
average lift; it uses every week of each engineer's record rather than a single
intervention week, accommodating adoption that is staggered across engineers and
time; and it draws no comparison between adopters and non-adopters, so unobserved
differences between those groups cannot bias it.

\paragraph{Dataset}
All three within-person analyses run on a common adopter panel:
\company{} engineers who recorded any \copilotcli{} or
\claudecode{} activity during \TreatedEverWindow{}, with the outcome aggregated
as merged PRs per engineer per week. The panel
\emph{includes} engineers who used both tools, since the dose-response and
demographic analyses pool across tools. We enforce single-tool purity in
one place --- the tool comparison in Section~\ref{sec:rq4} --- where attributing a
difference to one tool requires excluding engineers who used both. As in Part~1,
the panel is restricted to active engineers (at least two merged PRs in a
four-week pre-rollout window), and we apply a P5--P95 PR filter on pre-period
PR counts; the pre-period runs over the same dates as Part~1.

\subsubsection{Dose-response: does throughput rise with use? (RQ3)}
\label{sec:rq3-dose}

Developers do not benefit uniformly from AI tooling: the gain scales with how
intensively the tool is used~\citep{heilman2026ghcp}. We surface that variation
with a within-person dose-response on the dataset.

\paragraph{Design}
We fit on the dataset
\begin{equation}
  \log E\!\left[\text{PRs}_{i,w}\right]
  = \alpha_i + \tau_w + \sum_{k=1}^{4} \beta_k \, \mathbb{1}[d_{i,w} = k]
  + \beta_{5+} \, \mathbb{1}[d_{i,w} \geq 5],
  \label{eq:dose}
\end{equation}
where $d_{i,w}$ is engineer $i$'s tool-use-days in week $w$, binned into the
categories $\{0, 1, 2, 3, 4, 5{+}\}$ (the top bin, $5{+}$, collects weeks with
five or more tool-use days), $\alpha_i$ are engineer fixed
effects, $\tau_w$ are week fixed effects, and the reference category is
$d_{i,w} = 0$. Standard errors cluster on engineer. The design
follows the within-person template developed in Heilman and colleagues~\citep{heilman2026ghcp}.

Engineer fixed effects absorb every time-invariant engineer trait (skill, role,
team, project type, coding style); week fixed effects absorb every org-wide
weekly shock (holidays, all-hands, reorganizations, deadlines).

\begin{figure}[!htbp]
  \centering
  \includegraphics[width=0.75\columnwidth]{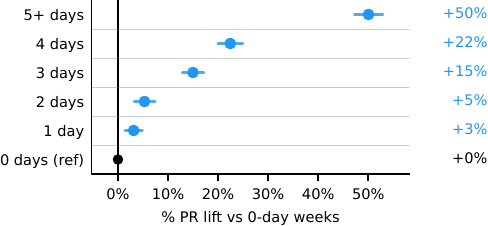}
  \caption{Percent change in merged PRs per engineer-week by days of tool use that
  week, versus a zero-day week.}
  \label{fig:dose}
\end{figure}

\paragraph{Results}
Figure~\ref{fig:dose} traces how many more PRs an engineer merges in weeks they use
the \toolclass{}s more. Each point is the lift in merged PRs for a
given number of tool-use days, relative to that same engineer's zero-day weeks. The
curve is monotone and well-separated, rising from \textbf{\DoseThreeDay{}\%} at three
days a week to \textbf{\DoseFiveDayPlus{}\%} at five or more.

Beyond raw volume, this dose-response may reflect a shift in \emph{what} work
developers take on. Several said the tools let them tackle tasks they would have
skipped---one said that, after the workshop, they made changes with ``nearly
100\% of the code written by Copilot CLI agent'' and that the tool gave them
``the ability to make larger changes that I never would have taken on in the
past, like splitting our huge test project into five separate files.'' Others
pointed to automating the tedious: ``boilerplate plumbing a lot faster,''
``repetitive unit tests faster,'' and prototypes ``much faster.'' Still others
emphasized parallel streams: ``I can be working on multiple things at once:
updating our documentation, analyzing it for issues and quality, prototyping app
ideas and code samples, creating tools for our team and others to use, having it
take care of busy work. So much more productive and unblocked.''

\subsubsection{RQ4 --- Does the specific tool matter?}
\label{sec:rq4}

\paragraph{Design}
Same fixed-effects Poisson framework as Equation~\ref{eq:dose}, but the dose
buckets are replaced by binary indicators for ``any \copilotcli{} use this
week'' and ``any \claudecode{} use this week.'' The contrast of interest is the
difference between the two coefficients, tested with a Wald statistic on the
engineer-clustered variance. We fit this on the single-tool subset defined above.

\begin{figure}[!htbp]
  \centering
  \includegraphics[width=0.75\columnwidth]{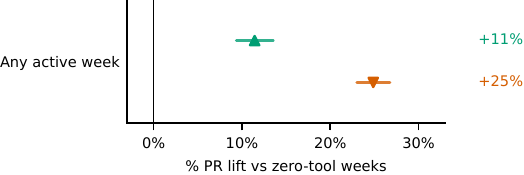}
  \caption{Percent change in merged PRs per engineer-week for
  \claudecode{}~\claudemarker{} and \copilotcli{}~\copilotmarker{}.}
  \label{fig:rq4}
\end{figure}

\paragraph{Results}
Figure~\ref{fig:rq4} compares PR lifts for adopters of each \toolclass{}. Each
marker is the lift in merged PRs in weeks an engineer used that
tool, relative to their own zero-tool weeks:

\begin{itemize}
  \item \claudecode{}~\claudemarker{}, any use: {\ToolClaudeLift{}\%} [\ToolClaudeLiftLo{}\%, \ToolClaudeLiftHi{}\%]
  \item \copilotcli{}~\copilotmarker{}, any use: {\ToolCopilotLift{}\%} [\ToolCopilotLiftLo{}\%, \ToolCopilotLiftHi{}\%]
\end{itemize}

\noindent
Overall, \copilotcli{} adopters experienced about a
\ToolRatio{}$\times$ the lift of \claudecode{} adopters (\ToolContrastPText{}),
comparing any-use weeks against the same engineer's zero-tool weeks.

The survey does not explain why \copilotcli{} outpaces \claudecode{}, but
it hints at shifting perceptions. Developers described both tools as
``impressively good and only getting better,'' and several reported migrating
toward \copilotcli{} after the workshop---one ``left [Claude Code] behind and
been using exclusively [\copilotcli{}].''

\subsubsection{RQ5 --- Among engineers who use these tools, who benefits most?}
\label{sec:rq5}

\paragraph{Design}
Same FE Poisson framework as Equation~\ref{eq:dose}, with the dose linearized
as days/week and interacted with a demographic indicator (career stage or
tenure).
In particular, career stage is a single HR snapshot, so promotions during the post-period are not captured.
We report the lift evaluated at 3 days/week for each subgroup.
These subgroup models specify usage as a continuous, constant-per-day slope
rather than the non-parametric buckets of the pooled dose-response in
Figure~\ref{fig:dose}. Both evaluate the same point---three days of use per
week---but because the curve in Figure~\ref{fig:dose} is convex, the linear
estimate sits somewhat above the corresponding three-day bucket. The subgroup
values are therefore best compared with one another rather than read directly
against the bucketed figure.
We control the false discovery rate across multiple tests using the
Benjamini--Hochberg procedure within each
moderator family (career, tenure) at $\alpha = 0.05$ on the interaction
$p$-values.
We correct here because RQ5 enumerates many subgroups --- the most subgroup
tests of any of our questions (eight career levels and five tenure bands) ---
making it the most exposed to a chance finding; the pre-specified, reference-anchored
contrasts elsewhere test far fewer hypotheses and do not pose the same multiplicity risk.
The design absorbs the same confounders as the dose model
(Section~\ref{sec:rq3-dose}); the interaction tells us whether the within-engineer
association varies by subgroup, but does not identify whether subgroups
differ because of tool affordances or because of subgroup-specific task mix.

\paragraph{Career stage}
\label{sec:career}

\begin{figure}[!htbp]
  \centering
  \includegraphics[width=0.75\columnwidth]{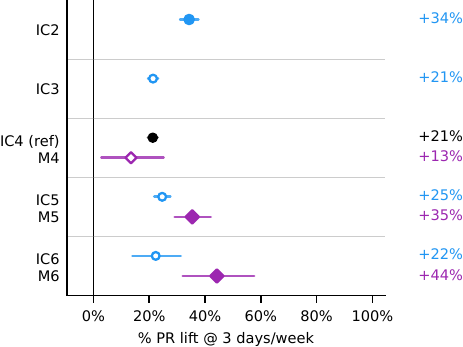}
  \caption{Percent PR lift at 3 days/week for individual contributors~\icmarker{}
  and managers~\mmarker{}, versus a mid-level individual contributor (IC4).}
  \label{fig:career}
\end{figure}

Figure~\ref{fig:career} shows how the PR lift varies by career stage. Each marker is a
subgroup's lift at three tool-use days per week,
relative to that subgroup's own zero-tool weeks. IC4 reference
adopters experienced a
\textbf{\CareerIcFourLift{}\%} [\CareerIcFourLiftLo{}\%, \CareerIcFourLiftHi{}\%] lift
in the number of PRs they created.
Compared to this reference,
the figure indicates that more junior ICs and more senior managers experience a larger PR lift.

\paragraph{Tenure}
\label{sec:tenure}

\begin{figure}[!htbp]
  \centering
  \includegraphics[width=0.75\columnwidth]{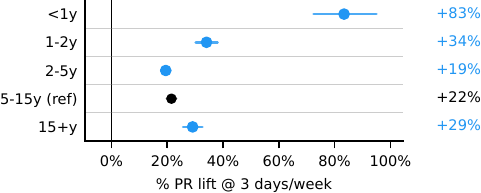}
  \caption{Percent PR lift at 3 days/week by tenure, versus an engineer who had been
  at \company{} for 5--15 years.}
  \label{fig:tenure}
\end{figure}

Figure~\ref{fig:tenure} shows how the PR lift varies with tenure at \company{}. As
with career stage, each marker is a tenure
band's lift at three tool-use days per week, relative to that band's own zero-tool
weeks. We restricted this analysis to engineers employed $\geq$6 months at first
tool-use, to mitigate the onboarding ramp-up confound --- newer hires would naturally
show a rise in merged pull requests independent of tool use. For the reference band,
engineers of 5--15 years' tenure experienced a \textbf{\TenureFiveFifteenLift{}\%}
[\TenureFiveFifteenLiftLo{}\%, \TenureFiveFifteenLiftHi{}\%] lift. Again we observe a ``C'' shape, where low- and high-tenure engineers
experienced larger lifts than mid-tenure engineers.

It is worth noting that the $<$1y point estimate is large; even halved, it would
remain among the largest moderator effects we observe in the study. We read this
large effect with caution --- despite trying to mitigate ramp-up
curves by excluding engineers with $<$6 months of tenure, those in this $<$1y bucket may
still be onboarding, in which case these estimates remain conflated
with that onboarding.

Across these themes, the PR gains look like more than faster typing or faster
code generation. The survey accounts instead point to changes upstream of the
PR: developers taking on different kinds of work, running parallel streams of
activity, and becoming more willing to attempt complex or previously deferred
tasks.

\section{Threats to Validity}
\label{sec:threats}

\emph{Construct validity.}
In the adoption study,
two thresholds are somewhat arbitrary: retention (5 of 14 days from first use) and
the 28-day merge window. We re-ran retention at both a looser 3-of-14 and a
stricter 7-of-14 threshold, and the retention model fits at all three. The 28-day
merge window trades longer-merging PRs for comparability and freedom from
right-censoring; much longer windows would be needed to gauge multi-year adoption.

\emph{Internal validity.} The adoption study is cross-sectional, so it cannot rule out
several confounds. For the social effect, we cannot separate peer influence from
homophily: an engineer may adopt because a peer did, or simply because similar
engineers tend to cluster together~\cite{aral2009distinguishing}. Our fixed
effects also miss engineer-specific, time-varying shocks (a personal project, a
vacation) that can move initial use within a single week. The outcomes study's
within-person design addresses this for outcomes, but no analogue exists for the
initial-use choice. We excluded retracted \claudecode{} licensees and two
broadly-licensed divisions from the adoption sample; re-running the fits on
two sample-frame variants left direction and rank-order unchanged --- top-bucket
reviewer-peer exposure, for example, stayed within a few points of its
\RevPeerInitHigh{}\% headline odds lift.

Merged PRs are an imperfect proxy for throughput and reward small, frequent PRs;
we may also miss quality costs such
as added complexity. In Part~1, adopters self-select, so the synthetic
counterfactual may be biased; re-running it with a placebo intervention at \PlaceboDate{} returned
\CIPlaceboLift{}\% [\CIPlaceboLiftLo{}\%, \CIPlaceboLiftHi{}\%], passing this check. In
Part~2, heavier-use weeks may carry a lighter task mix, so we read the
dose-response as an association, not a cause; Heilman and colleagues
controlled for effort and still found gains~\citep{heilman2026ghcp},
mitigating the concern somewhat. The
tool comparison's single-tool restriction also excludes possibly-distinct
multi-tool users.

\emph{External validity.} Both studies draw on one company in a single early-2026 window, so findings
generalize only insofar as other settings resemble \company{}'s. That said,
\company{} is a large, globally distributed software company whose engineers span
many product domains, work across diverse codebases, and program in a wide range
of languages, which lends some breadth to the setting. We also
measured PRs from Azure DevOps only, undercounting engineers on other ecosystems
in both studies.

Our 16-week window is long enough to surface an early fade
but not a long-horizon decline.

\emph{Researcher positionality.} The authors are \company{} employees; \company{} sells AI tools, encourages their
use, and owns GitHub, the maker of \copilotcli{}. Though under no explicit
pressure, our proximity may have shaped our questions, design, and
interpretation.

\section{Discussion}
\label{sec:discussion}

\textbf{Adoption is substantially social.} The strongest predictor of who tries
\copilotcli{} in any given week is whether the engineer's peers --- especially
the broader skip-level group --- have already tried it.
These social signals could reflect peers observing each other's successful
work strategies~\cite{xiao2014social}, or, in the case of managers' use,
implicit or explicit directives.
To the extent that our results reflect prior work --- that learning developer
tools from peers is highly effective~\cite{murphy2011peer,murphy2015users} ---
this suggests that organizations should enable engineers' use
of \toolclass{}s to be visible and socially reinforced.

An engineer's prior IDE Copilot use also predicts trying \copilotcli{};
we interpret this as an indicator of familiarity and openness to AI tooling.

\textbf{Retention is substantially behavioral.}
The predictors associated with adoption are not equally associated with
retention, and one of them --- prior IDE Copilot use --- actively predicts
\emph{against} sticking with \copilotcli{}.

A plausible interpretation is that engineers who already trust AI tooling in
their IDE will try \copilotcli{} but have a familiar alternative
to fall back on, so they may not build a sustained \copilotcli{} habit; engineers
for whom \copilotcli{} is their first such tool have no such fallback and, if they
stay, tend to stay more firmly.

\textbf{Engineer attributes do not matter much to adoption and retention.}
Career stage produces a
gentle gradient among individual contributors and nothing detectable among managers. Tenure produces
a small newest-hire bump and otherwise nothing. The five groups together suggest
that \emph{what} an engineer does (peer ties, prior tool use, PR
cadence) explains who adopts and retains far better than \emph{who} an engineer
is in the organization.

\textbf{Pull request lifts persist.}
Our finding that the PR lift does not fade stands
in contrast to He and colleagues~\citep{he2026cursor}, whose Cursor lift faded at
month two and was gone by month three --- well inside our roughly four-month
window, so a too-short observation period is an unlikely explanation. We read the
contrast as genuine, and offer two non-mutually-exclusive reasons:
(a) \textit{tool generation} --- they study Cursor, a 2024-2025 IDE-based tool, while
we study 2026 \toolclass{}s; and
(b) \textit{unit of analysis} --- their repo-level average reverts toward the
mean as adoption spreads from keen early adopters to marginal users, whereas our
within-person design conditions on the same engineer and is immune to this
compositional drift.

\textbf{Why does \copilotcli{} outpace \claudecode{}?}
The larger PR lift for \copilotcli{} than \claudecode{} adopters
(Section~\ref{sec:rq4}) is surprising in light of developer sentiment around
\toolclass{}.
As described in the introduction, public early-2026 comparisons of agentic
coding tools generally rate \claudecode{} as the preferred option for autonomous agentic work; our data
shows the opposite ordering on merged-PR throughput.
We have two hypotheses:
(a) the two tools are used for \textit{different task mixes} --- engineers reach
for the two tools for different purposes; and
(b) \textit{\copilotcli{} works better for \company{} employees} --- because
\company{} owns GitHub, the maker of \copilotcli{}, but is a buyer of
\claudecode{}, organizational forces likely helped align the \copilotcli{}
harness with the way engineers at \company{} work.

\section{Conclusion}
\label{sec:conclusion}

In this paper, we analyzed how emerging CLI-based agentic coding tools are
adopted and used in a large software organization.
We found that initial use spread substantially through social channels --- an
engineer's peers and managers using the tool --- and that adopters went on to
merge roughly \AbstractPRLift{}\% more pull requests, a gain that held steady
across our four-month window.
For organizations weighing the token costs raised at the outset, this sustained
lift in merged pull requests is direct evidence that \toolclass{}s can move a
concrete output metric --- though whether that output justifies its cost is a
question of value, not throughput alone.
The pressing open question is now about quality, whether this added
throughput yields better software.
The field still lacks agreed-upon measures to answer
it, and building them should be a priority of the research community.
While AI developer tools are advancing at a rapid pace --- from chat to code completion to
agents on the desktop and in the cloud --- understanding adoption and outcomes
is critical for today's software organizations to
align expectations with reality.

\section*{Acknowledgments}

Thanks to the many Microsoft employees who participated in surveys and experimented with the tools.
Thanks also to the teams who collect and maintain data that supports this research.
Special thanks to
Barrett Amos,
Alex Heilman,
Alex Kyllo,
David Spiers,
and many others for their help with data access, analysis, and interpretation.

In preparing this work the authors used GitHub Copilot (predominantly Anthropic
Claude Opus models, versions 4.6--4.8) for three purposes: to draft an initial
version of the manuscript from the underlying analysis notebooks; to implement
the data-analysis and figure-generating code that produces every quantitative
result and figure in Sections~\ref{sec:study1} and~\ref{sec:study2}; and to
assist with subsequent reorganization and copy-editing of the text at the
authors' direction. All research questions, study designs, methodological
choices, and interpretations are the authors' own. The authors take full responsibility
for the content, accuracy, and conclusions of this work.

\bibliographystyle{plainnat}
\bibliography{references}

@inproceedings{he2026cursor,
  title={Speed at the Cost of Quality: How {Cursor} {AI} Increases Short-Term Velocity and Long-Term Complexity in Open-Source Projects},
  author={He, Hao and Miller, Courtney and Agarwal, Shyam and K{\"a}stner, Christian and Vasilescu, Bogdan},
  booktitle={Proceedings of the 23rd International Conference on Mining Software Repositories (MSR)},
  year={2026},
}

@inproceedings{agarwal2026agentic,
  title={{AI} {IDEs} or Autonomous Agents? Measuring the Impact of Coding Agents on Software Development},
  author={Agarwal, Shyam and He, Hao and Vasilescu, Bogdan},
  booktitle={Proceedings of the 23rd International Conference on Mining Software Repositories (MSR), Mining Challenge Track},
  year={2026},
}

@inproceedings{paradis2025much,
  title={How much does {AI} impact development speed? An enterprise-based randomized controlled trial},
  author={Paradis, Elise and Grey, Kate and Madison, Quinn and Nam, Daye and Macvean, Andrew and Meimand, Vahid and Zhang, Nan and Ferrari-Church, Ben and Chandra, Satish},
  booktitle={2025 IEEE/ACM 47th International Conference on Software Engineering: Software Engineering in Practice (ICSE-SEIP)},
  pages={618--629},
  year={2025},
  organization={IEEE}
}

@inproceedings{vaithilingam2022expectation,
  title={Expectation vs. experience: Evaluating the usability of code generation tools powered by large language models},
  author={Vaithilingam, Priyan and Zhang, Tianyi and Glassman, Elena L},
  booktitle={{CHI} conference on human factors in computing systems extended abstracts},
  pages={1--7},
  year={2022}
}

@misc{pragmaticengineer2026,
  author       = {Orosz, Gergely},
  title        = {{AI} {Tooling} for {Software} {Engineers} in 2026},
  howpublished = {\url{https://newsletter.pragmaticengineer.com/p/ai-tooling-2026}},
  year         = {2026},
  note         = {Accessed 2026-05-20}
}

@misc{stackoverflow2025,
  author       = {{Stack Overflow}},
  title        = {Developers remain willing but reluctant to use {AI}: The 2025 Developer Survey results are here},
  howpublished = {\url{https://stackoverflow.blog/2025/12/29/developers-remain-willing-but-reluctant-to-use-ai-the-2025-developer-survey-results-are-here/}},
  year         = {2025},
  month        = dec,
  note         = {Accessed 2026-05-20}
}

@misc{fortune2026meta,
  author       = {{Fortune}},
  title        = {A {Meta} employee created a dashboard so coworkers can compete to be the company's No. 1 {AI} token user -- and {Zuckerberg} doesn't even rank in the top 250},
  howpublished = {\url{https://fortune.com/2026/04/09/meta-killed-employee-ai-token-dashboard/}},
  year         = {2026},
  month        = apr,
  note         = {Accessed 2026-05-20}
}

@misc{heilman2026ghcp,
  title={{GitHub Copilot} and Developer Productivity: An Observational Dose-Response Analysis},
  author={Heilman, Alex and Kyllo, Alex and Murphy-Hill, Emerson},
  journal={arXiv preprint arXiv:2606.00438},
  year={2026}
}

@article{brodersen2015causalimpact,
  title={Inferring causal impact using {Bayesian} structural time-series models},
  author={Brodersen, Kay H and Gallusser, Fabian and Koehler, Jim and Remy, Nicolas and Scott, Steven L},
  journal={Annals of Applied Statistics},
  volume={9},
  pages={247--274},
  year={2015}
}

@article{bhattacherjee2001continuance,
  title={Understanding Information Systems Continuance: An Expectation-Confirmation Model},
  author={Bhattacherjee, Anol},
  journal={MIS quarterly},
  volume={25},
  number={3},
  pages={351--370},
  year={2001},
  publisher={Management Information Systems Research Center, University of Minnesota}
}

@book{rogers2003diffusion,
  title={Diffusion of Innovations, 5th Edition},
  author={Rogers, Everett M.},
  isbn={9780743258234},
  lccn={2003049022},
  year={2003},
  publisher={Simon and Schuster}
}

@article{davis1989perceived,
  title={Perceived usefulness, perceived ease of use, and user acceptance of information technology},
  author={Davis, Fred D},
  journal={MIS quarterly},
  volume={13},
  number={3},
  pages={319--340},
  year={1989},
  publisher={Management Information Systems Research Center, University of Minnesota}
}

@article{venkatesh2000theoretical,
  title={A theoretical extension of the technology acceptance model: Four longitudinal field studies},
  author={Venkatesh, Viswanath and Davis, Fred D},
  journal={Management science},
  volume={46},
  number={2},
  pages={186--204},
  year={2000},
  publisher={INFORMS}
}

@article{venkatesh2003user,
  title={User acceptance of information technology: Toward a unified view1},
  author={Venkatesh, Viswanath and Morris, Michael G and Davis, Gordon B and Davis, Fred D},
  journal={MIS quarterly},
  volume={27},
  number={3},
  pages={425--478},
  year={2003},
  publisher={Management Information Systems Research Center, University of Minnesota}
}

@inproceedings{xiao2014social,
  title={Social influences on secure development tool adoption: why security tools spread},
  author={Xiao, Shundan and Witschey, Jim and Murphy-Hill, Emerson},
  booktitle={Proceedings of the 17th ACM conference on Computer supported cooperative work \& social computing},
  pages={1095--1106},
  year={2014}
}

@article{reyes2026adoption,
  title={Adoption of AI tools in software development: a systematic literature review},
  author={Reyes-Reina, Dar{\'\i}o and Sanch{\'e}z-Torres, Jenny Marcela and Rueda-C{\'a}ceres, Iv{\'a}n Mauricio},
  journal={Science of Computer Programming},
  volume={254},
  pages={103521},
  year={2026},
  publisher={Elsevier}
}

@article{daniotti2026using,
  title={Who is using AI to code? Global diffusion and impact of generative AI},
  author={Daniotti, Simone and Wachs, Johannes and Feng, Xiangnan and Neffke, Frank},
  journal={Science},
  pages={eadz9311},
  year={2026},
  publisher={American Association for the Advancement of Science}
}

@article{robbes2026agentic,
  title={Agentic Much? Adoption of Coding Agents on GitHub},
  author={Robbes, Romain and Matricon, Th{\'e}o and Degueule, Thomas and Hora, Andre and Zacchiroli, Stefano},
  journal={arXiv preprint arXiv:2601.18341},
  year={2026}
}

@article{yang6803624claude,
  title={Claude Code Scientists: Measuring AI Adoption and Productivity among Scientists},
  author={Yang, Charles},
  journal={Available at SSRN 6803624},
  year={2026}
}

@Article{peng2023,
  author  = {Sida Peng and Eirini Kalliamvakou and Peter Cihon and Mert Demirer},
  title   = {The Impact of {AI} on Developer Productivity: Evidence from {GitHub Copilot}},
  journal = {arXiv preprint arXiv:2302.06590},
  year    = {2023}
}

@Article{cui2025,
  author  = {Zheyuan Cui and Mert Demirer and Sonia Jaffe and Leon Musolff and Sida Peng and Tobias Salz},
  title   = {The Effects of Generative {AI} on High-Skilled Work: Evidence from Three Field Experiments with Software Developers},
  journal = {Management Science},
  year    = {2025},
  note    = {Forthcoming},
  doi     = {10.2139/ssrn.4945566}
}

@article{mohamed2026,
author = {Mohamed, Amr and Assi, Maram and Guizani, Mariam},
title = {The Impact of LLM-Assistants on Software Developer Productivity: A Systematic Review and Mapping Study},
year = {2026},
publisher = {Association for Computing Machinery},
address = {New York, NY, USA},
issn = {1049-331X},
url = {https://doi.org/10.1145/3809494},
doi = {10.1145/3809494},
note = {Just Accepted},
journal = {ACM Trans. Softw. Eng. Methodol.},
month = apr
}

@inproceedings{ziegler2022productivity,
  title={Productivity assessment of neural code completion},
  author={Ziegler, Albert and Kalliamvakou, Eirini and Li, X Alice and Rice, Andrew and Rifkin, Devon and Simister, Shawn and Sittampalam, Ganesh and Aftandilian, Edward},
  booktitle={Proceedings of the 6th ACM SIGPLAN international symposium on machine programming},
  pages={21--29},
  year={2022}
}

@article{gurgul2026state,
  title={The State of Generative AI in Software Development: Insights from Literature and a Developer Survey},
  author={Gurgul, Vincent and Gubela, Robin and Lessmann, Stefan},
  journal={arXiv preprint arXiv:2603.16975},
  year={2026}
}

@article{o2025more,
  title={More code, less validation: Risk factors for over-reliance on AI coding tools among scientists},
  author={O'Brien, Gabrielle and Parker, Alexis and Eisty, Nasir and Carver, Jeffrey},
  journal={arXiv preprint arXiv:2512.19644},
  year={2025}
}

@misc{afroz2026,
      title={The Fast and Spurious: Developer Productivity with GenAI}, 
      author={Sadia Afroz and Zixuan Feng and Tyler Menezes and Katie Kimura and Bianca Trinkenreich and Igor Steinmacher and Anita Sarma},
      year={2026},
      eprint={2510.24265},
      archivePrefix={arXiv},
      primaryClass={cs.SE},
      url={https://arxiv.org/abs/2510.24265}, 
}

@inproceedings{mozannar2024reading,
  title={Reading between the lines: Modeling user behavior and costs in AI-assisted programming},
  author={Mozannar, Hussein and Bansal, Gagan and Fourney, Adam and Horvitz, Eric},
  booktitle={Proceedings of the 2024 CHI conference on human factors in computing systems},
  pages={1--16},
  year={2024}
}

@inproceedings{brandebusemeyer2025developers,
  title={Developers' Experience with Generative AI--First Insights from an Empirical Mixed-Methods Field Study},
  author={Brandebusemeyer, Charlotte and Schimmer, Tobias and Arnrich, Bert},
  booktitle={Proceedings of the International Conference on Software Engineering (ICSE), Software Engineering in Practice Track (SEIP)},
  year={2026}
}

@article{weber2024significant,
  title={Significant productivity gains through programming with large language models},
  author={Weber, Thomas and Brandmaier, Maximilian and Schmidt, Albrecht and Mayer, Sven},
  journal={Proceedings of the ACM on Human-Computer Interaction},
  volume={8},
  number={EICS},
  pages={1--29},
  year={2024},
  publisher={ACM New York, NY, USA}
}

@inproceedings{shihab2025effects,
  title={The Effects of GitHub Copilot on Computing Students' Programming Effectiveness, Efficiency, and Processes in Brownfield Coding Tasks},
  author={Shihab, Md Istiak Hossain and Hundhausen, Christopher and Tariq, Ahsun and Haque, Summit and Qiao, Yunhan and Mulanda, Brian Wise},
  booktitle={Proceedings of the 2025 ACM Conference on International Computing Education Research V. 1},
  pages={407--420},
  year={2025}
}

@inproceedings{kumar2025,
author = {Kumar, Aayush and Bajpai, Yasharth and Gulwani, Sumit and Soares, Gustavo and Murphy-Hill, Emerson},
title = {Why AI Agents Still Need You: Findings from Developer-Agent Collaborations in the Wild},
year = {2025},
publisher = {IEEE Press},
url = {https://doi.org/10.1109/ASE63991.2025.00043},
doi = {10.1109/ASE63991.2025.00043},
booktitle = {2025 40th IEEE/ACM International Conference on Automated Software Engineering (ASE)},
pages = {432–444},
numpages = {13},
location = {Seoul, Korea, Republic of}
}

@article{becker2025measuring,
  title={Measuring the impact of early-2025 AI on experienced open-source developer productivity},
  author={Becker, Joel and Rush, Nate and Barnes, Elizabeth and Rein, David},
  journal={arXiv preprint arXiv:2507.09089},
  year={2025}
}

@article{Cui2024Productivity,
	author = {Cui, Kevin Zheyuan and Demirer, Mert and Jaffe, Sonia and Musolff, Leon and Peng, Sida and Salz, Tobias},
	journal = {An MIT Exploration of Generative AI},
	year = {2024},
	month = {mar 27},
	note = {https://mit-genai.pubpub.org/pub/v5iixksv},
	publisher = {MIT},
	title = {The {Productivity} {Effects} of {Generative} {AI}: Evidence from a {Field} {Experiment} with {GitHub} {Copilot}},
}

@INPROCEEDINGS{butler2025,
  author={Butler, Jenna and Suh, Jina and Haniyur, Sankeerti and Hadley, Constance},
  booktitle={2025 IEEE/ACM 47th International Conference on Software Engineering: Software Engineering in Practice (ICSE-SEIP)}, 
  title={Dear Diary: A Randomized Controlled Trial of Generative AI Coding Tools in the Workplace}, 
  year={2025},
  volume={},
  number={},
  pages={319-329},
  doi={10.1109/ICSE-SEIP66354.2025.00034}}

@article{song2024impact,
  title={The impact of generative AI on collaborative open-source software development: Evidence from GitHub Copilot},
  author={Song, Fangchen and Agarwal, Ashish and Wen, Wen},
  journal={arXiv preprint arXiv:2410.02091},
  year={2024}
}

@article{yeverechyahu2024impact,
  title={The impact of large language models on open-source innovation: Evidence from github copilot},
  author={Yeverechyahu, Doron and Mayya, Raveesh and Oestreicher-Singer, Gal},
  journal={arXiv preprint arXiv:2409.08379},
  year={2024}
}

@article{kreitmeir2024heterogeneous,
  title={The heterogeneous productivity effects of generative ai},
  author={Kreitmeir, David and Raschky, Paul A},
  journal={arXiv preprint arXiv:2403.01964},
  year={2024}
}

@techreport{Demirer2026,
 title = "Writing Code vs. Shipping Code: Productivity Effects Across Generations of AI Coding Tools",
 author = "Demirer, Mert and Musolff, Leon and Yang, Liyuan",
 institution = "National Bureau of Economic Research",
 type = "Working Paper",
 series = "Working Paper Series",
 number = "35275",
 year = "2026",
 month = "May",
 doi = {10.3386/w35275},
 URL = "http://www.nber.org/papers/w35275",
}

@inproceedings{stray2026,
  title={Developer Productivity With and Without GitHub Copilot: A Longitudinal Mixed-Methods Case Study},
  author={Stray, Viktoria and Brandtzæg, Elias Goldmann and Wivestad, Viggo and Barbala, Astri and Moe, Nils Brede},
  booktitle={Proceedings of the 59th Hawaii International Conference on System Sciences},
  year={2026}
}

@article{kumar2025intuition,
  title={Intuition to Evidence: Measuring AI's True Impact on Developer Productivity},
  author={Kumar, Anand and Khare, Vishal and Sharma, Deepak and Kumar, Satyam and Saini, Vijay and Yadav, Anshul and Jain, Sachendra and Rana, Ankit and Verma, Pratham and Meena, Vaibhav and others},
  journal={arXiv preprint arXiv:2509.19708},
  year={2025}
}

@article{tomaz2026impacts,
  title={Impacts of Generative AI on Agile Teams' Productivity: A Multi-Case Longitudinal Study},
  author={Tomaz, Rafael and Guenes, Paloma and Ara{\~A}{\=e}jo, Allysson Allex and Baldassarre, Maria Teresa and Kalinowski, Marcos},
  journal={arXiv preprint arXiv:2602.13766},
  year={2026}
}

@article{quispe2026coding,
  title={Coding Beyond Your Training: Claude Code and the Technological Frontier of Software Developers},
  author={Quispe, Alexander},
  journal={arXiv preprint arXiv:2605.25438},
  year={2026}
}

@article{aral2009distinguishing,
  title={Distinguishing influence-based contagion from homophily-driven diffusion in dynamic networks},
  author={Aral, Sinan and Muchnik, Lev and Sundararajan, Arun},
  journal={Proceedings of the National Academy of Sciences},
  volume={106},
  number={51},
  pages={21544--21549},
  year={2009},
  publisher={National Academy of Sciences}
}

@article{murphy2015users,
  title={How do users discover new tools in software development and beyond?},
  author={Murphy-Hill, Emerson and Lee, Da Young and Murphy, Gail C and McGrenere, Joanna},
  journal={Computer Supported Cooperative Work (CSCW)},
  volume={24},
  number={5},
  pages={389--422},
  year={2015},
  publisher={Springer}
}

@inproceedings{murphy2011peer,
  title={Peer interaction effectively, yet infrequently, enables programmers to discover new tools},
  author={Murphy-Hill, Emerson and Murphy, Gail C},
  booktitle={Proceedings of the ACM 2011 conference on Computer supported cooperative work},
  pages={405--414},
  year={2011}
}

\end{document}